\documentclass[12pt, draftclsnofoot, onecolumn]{IEEEtran}
\usepackage{setspace}
\usepackage{cite}
\usepackage{amsmath,amssymb,amsfonts,bm}
\usepackage{amsthm}
\usepackage{algorithmic}
\usepackage{graphicx}
\usepackage{multirow}
\usepackage{float}
\usepackage{textcomp}
\usepackage{verbatim}
\usepackage[outdir=./]{epstopdf}
\usepackage{xcolor}
\usepackage{subcaption}
\usepackage{hhline}
\usepackage[ruled,vlined]{algorithm2e}

\def\BibTeX{{\rm B\kern-.05em{\sc i\kern-.025em b}\kern-.08em
    T\kern-.1667em\lower.7ex\hbox{E}\kern-.125emX}}
\title{Spatial Modulation with Energy Detection: Diversity Analysis and Experimental Evaluation}

\author{$\text{Elio Faddoul}$, \textit{Graduate Student Member, IEEE}, $\text{Ghassan M. Kraidy}$, \textit{Senior Member, IEEE}, $\text{Constantinos Psomas}$, \textit{Senior Member, IEEE}, $\text{Symeon Chatzinotas}$, \textit{Fellow IEEE}, and $\text{Ioannis Krikidis}$, \textit{Fellow, IEEE} \thanks{This  work received funding from the European Research Council (ERC) under the European Union's Horizon 2020 research and innovation programme (Grant agreement No. 819819). This work was also co-funded by the European Regional Development Fund and the Republic of Cyprus through the Research and Innovation Foundation, under the project CONCEPT/0722/0123 (FLOAT).} \thanks{E. Faddoul, C. Psomas, and I. Krikidis are with the IRIDA Research Centre for Communication Technologies, Department of Electrical and Computer Engineering, University of Cyprus, Cyprus (e-mail: \{efaddo01, psomas, krikidis\}@ucy.ac.cy). G.M. Kraidy is with the Department of Electronic Systems, Norwegian University of Science and Technology, Gj{\o}vik, Norway (e-mail: {ghassan.kraidy@ntnu.no}). S. Chatzinotas is with the Interdisciplinary Centre for Security, Reliability and Trust (SnT), University of Luxembourg, L-1855, Luxembourg (email: symeon.chatzinotas@uni.lu)}\thanks{Part of this work was presented at the IEEE
Global Communications Conference, Rio de Janeiro, Brazil, December 2022 \cite{Faddoul2022low}.}}
\begin{document}
\maketitle
\vspace{-17mm}
\begin{abstract}
\vspace{-3mm}
In this paper, we present a non-coherent energy detection scheme for spatial modulation (SM) systems. In particular, the use of SM is motivated by its low-complexity implementation in comparison to multiple-input multiple-output (MIMO) systems, achieved through the activation of a single antenna during transmission. Moreover, energy detection-based communications restrict the channel state information to the magnitude of the fading gains. This consideration makes the design applicable for low-cost low-powered devices since phase estimation and its associated circuitry are avoided. We derive an energy detection metric for a multi-antenna receiver based on the maximum-likelihood (ML) criterion. By considering a biased pulse amplitude modulation, we develop an analytical framework for the SM symbol error rate at high signal-to-noise ratios. Numerical results show that the diversity order is proportional to half the number of receive antennas; this result stems from having partial receiver channel knowledge. In addition, we compare the performance of the proposed scheme with that of the coherent ML receiver and show that the SM energy detector outperforms its coherent counterpart in certain scenarios, particularly when utilizing non-negative constellations. Ultimately, we implement an SM testbed using software-defined radio devices and provide experimental error rate measurements that validate our theoretical contribution.
\end{abstract}
\vspace{-3mm}
\begin{IEEEkeywords}
\vspace{-3mm}
Spatial modulation, energy detection, biased pulse-amplitude modulation, diversity, software-defined radio, experimental results.
\end{IEEEkeywords}
\section{Introduction}
Motivated by the recent growth in the number of connected devices and the evolution of Internet-of-Things (IoT) applications \cite{al2015internet}, future wireless communication systems require an intelligent utilization of the available resources. Specifically, due to the unparalleled increase in data traffic, new transmission technologies have been studied to minimize transceiver complexity and cost. A strong candidate that addresses this challenge is index modulation (IM), which was initially introduced in the framework of spatial modulation (SM) for multiple-input multiple-output (MIMO) systems \cite{mesleh2008spatial}. In the context of SM, alongside the information carried by a modulation symbol, additional information bits are transmitted by activating a single antenna index at the transmitter. This technology has proven to achieve high spectral and energy efficiency while maintaining a simple design architecture in comparison to traditional MIMO and single-antenna systems \cite{mesleh2008spatial}, \cite{wen2019survey}. At the receiver side, SM detection is twofold, i.e., the detection of the active transmit antenna index as well as the detection of the data symbol transmitted over the activated transmit antenna. Since only one antenna is activated, a single radio-frequency (RF) chain is sufficient for transmission, hence significantly reducing energy consumption. Consequently, SM systems do not require synchronization among the transmitting antennas and avoid the problem of inter-channel interference which typically arises when multiple antennas are activated \cite{wen2019survey},\cite{stavridis2013energy}.

The availability of channel state information (CSI) holds significant importance in signal design and detection \cite{feng2012survey}. The techniques for detecting signals can be broadly classified into two categories based on the availability of CSI: coherent detection and non-coherent detection. Coherent detection is a method that depends on the receiver's ability to estimate the CSI accurately. This detection technique is appropriate for systems with adequate resources for CSI estimation and a sufficiently long coherence time so that the CSI can be exploited before it changes. In contrast, non-coherent detection refers to scenarios where obtaining accurate CSI is not practical and hence relies on statistical CSI obtained by averaging the channel response over an extended period. However, this process may lead to undesirable delays in the communication system \cite{al2019error}. On the other hand, the deployment of large-scale IoT networks requires low-cost, low-power, and low-complexity devices. Accordingly, energy detection-based non-coherent systems have been exploited to meet the requirements of the IoT due to their capability of operating with partial receiver CSI knowledge \cite{gao2018energy}. In other words, only channel magnitude estimates are available at the receiver whereas phase estimates are ignored. Inherently, power-intensive circuitry, such as oscillators and mixers is avoided \cite{mallik2017channel}. However, this decrease in complexity and power consumption comes at the cost of sub-optimal performance compared with coherent systems. Thus, a tradeoff between complexity and performance is introduced.

Another advantage that supports the use of energy detection is in terms of channel estimation. More specifically, the complexity involved in estimating the channel for a non-coherent energy detector is comparatively lower than that of a coherent detector. The former requires the estimation of only the channel magnitudes, whereas the latter needs to estimate the complete channel matrix \cite{li2019channel}. Moreover, since the receiver incorporates a form of square-law detection, the symbol energy levels used must be chosen to limit the symbol error probability. A popular constellation design consists of non-negative (biased) signaling, such as biased pulse amplitude modulation (PAM)\cite{gao2018energy}. The use of non-coherent systems with biased PAM constellations has been documented for the emerging $60$-GHz millimeter-wave (mmWave) wireless personal area networks based on the ECMA-$387$ standard, specifically for devices with low power consumption \cite[Sec. 1.5.2]{yong201160ghz}. In this context, non-coherent energy detection has been proven to achieve a better rate-energy tradeoff than its coherent counterpart for systems with energy harvesting modules, especially when incorporating an integrated architecture of energy harvesters and information decoders for simultaneous wireless information and power transfer (SWIPT)\cite{park2020transmitter}.

Various detection schemes corresponding to different complexity levels have been explored for SM. For instance, the maximum-likelihood (ML) detector, which characterizes joint detection and searches over all possible transmit antennas and constellation symbols was first introduced in \cite{jeganathan2008spatial}. The results showed that by employing an ML detector, significant performance gains (around 4 dB) can be achieved as opposed to the scheme in \cite{mesleh2008spatial}. In contrast, to mitigate the substantial search complexity inherent in ML detection, numerous studies have explored alternative suboptimal detection algorithms, which effectively reduce computational overhead. Notable among these algorithms are signal vector-based detection \cite{wang2011signal}, maximal ratio combining (MRC) \cite{maleki2016mrc}, and sphere decoding detection \cite{al2019optimum}. Nevertheless, these solutions consider optimal and sub-optimal architectures with full receiver CSI knowledge, which in some cases cannot be perfectly obtained, especially for low-complexity devices. Concerning non-coherent communication, several techniques have been studied for SM systems from a differential encoding/decoding standpoint\cite{ishikawa201850}. Such detection schemes are based on observation of the received signal for more than a single symbol interval, where a high level of synchronization accuracy is needed, as well as requiring additional bandwidth to transmit the difference between consecutive symbols, which can reduce the overall spectral efficiency \cite{bian2014differential}. More recently, SM techniques have witnessed integration with other emerging communication systems and applications. Notably, SM has been invoked in reconfigurable intelligent surfaces (RIS)-based systems to enhance the spectral efficiency of RIS-assisted wireless transmissions \cite{canbilen2022performance},\cite{sanila2023joint}. The main idea is to exploit both transmit and receive antenna indices to convey additional information by appropriately tuning the reflection coefficients of the passive RIS elements. Furthermore, other instances of SM integration include its application within the context of the IoT \cite{niu2023unified} and full-duplex radio systems \cite{liu2021transmit}.



Despite the rigorous theoretical contributions of SM systems, the experimental validations are limited. For instance, the first practical implementation of SM is demonstrated in \cite{younis2013performance} in an indoor environment. The work in \cite{serafimovski2013practical} collected realistic channel measurements for correlated and uncorrelated Rayleigh fading channels to analyze the performance of SM. More recently, the authors in \cite{hiari2020first} implemented the first space shift keying (SSK) system under Rician channel conditions. However, the detection in SSK is simpler than in SM since no actual constellation symbols are being transmitted. An experimental study of SM with a reconfigurable antenna in which radiation patterns act as the index for modulation has been implemented in \cite{kokar2019demo}. The work in \cite{zhou2022experimental} investigated the performance of SM systems with an experimental testbed in mixed line-of-sight/non-line-of-sight environments. In this regard, and to the best of our knowledge, the realization of a proper energy detection scheme for SM, which exploits the energy of both the channel coefficients and the information symbol, has yet to be accomplished. Therefore, in this work, we explore a novel approach that combines SM techniques at the transmitter with non-coherent energy detection at the receiver. Particularly, by considering a biased modulation specifically designed for energy detection, we analyze the performance of the proposed scheme in terms of error rate. Furthermore, we examine the performance of the SM energy detection scheme by establishing a simple experimental prototype to validate our analysis. Accordingly, the contributions of this paper are threefold:
\begin{itemize}
\item We develop a low-complexity SM energy detector equipped with multiple receive antennas, relying solely on knowledge of the channel magnitudes and energy of the symbols. The performance of the proposed energy detection scheme is analytically studied, whereby a  non-negative real constellation is considered. Owing to the complexity of the problem, we provide an accurate analytical framework for the computation of the overall error rate for SM at high signal-to-noise ratios (SNRs) for an arbitrary number of transmit and receive antennas by considering union-bound methods.
\item A thorough diversity analysis over slow-fading channels is provided. It is shown that the diversity order exhibited by the SM energy detector corresponds to half the number of receive antennas. This loss in diversity is due to the absence of full channel knowledge and the fact that the receiver relies on energy detection. Numerical results reveal that the proposed SM energy detector outperforms its coherent counterpart for specific scenarios where the receiver is equipped with double the number of receive antennas, and a biased modulation is used.
\item Finally, we implement an experimental SM testbed using software-defined radio (SDR) hardware tools for the proposed energy detection scheme with a biased modulation design, for the purpose of verifying our simulation and analytical results concerning the error rate performance. We show that the experimental error rate performance follows a similar trend as the simulation results, thereby validating our theoretical contributions. As a result of its low complexity, we conclude that the SM energy detector is appropriate for low-cost low-powered devices.
\end{itemize}

The paper is organized as follows. In Section II, we present the system model for the non-coherent SM energy detection scheme. In Section III, we derive analytical error rate expressions for the SM energy detector and provide an analysis regarding its diversity order. In Section V, we describe the experimental setup for the SM system with energy detection. Section VI provides the experimental and numerical results, and Section VII gives the concluding remarks.

\textit{Notation:} Lower and upper case boldface letters denote vectors and matrices, respectively; $[\cdot]^T$ denotes the transpose; $\boldsymbol{0}_{N}$ and ${\bf{I}}_{N}$ are the $N \times 1$ zero-vector and $N \times N$ identity matrix, respectively; $|\cdot|$ and $|\cdot|^2$ denote the absolute value and square absolute value, respectively; $\mathbb{P}\{ \cdot \}$ and $\mathbb{E}\{ \cdot\}$ represent the probability and expectation operators, respectively; $\binom{\cdot}{\cdot}$ denotes the binomial coefficient; $I_k(\cdot)$ denotes the modified Bessel function of the $k$-th order and the first kind; $\Gamma(\cdot)$ and $\gamma(\cdot,\cdot)$ are the complete and lower incomplete gamma function, respectively; $Q_{m}\left(\cdot,\cdot\right)$ denotes the generalized Marcum-$Q$ function of order $m$ and $_2F_1\left(\cdot,\cdot;\cdot;\cdot\right)$ is the Gaussian hypergeometric function.
\section{System Model}
\subsection{Topology}
\begin{figure*}[!t]
\includegraphics[width=\textwidth]{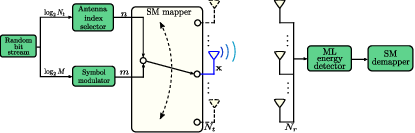}
\centering
\caption{Transceiver structure for the energy detector-based non-coherent SM system.}\label{fig:1}
\end{figure*}
We consider an $N_t \times N_r$ SM system with $N_t$ transmit antennas and $N_r$ receive antennas. At the transmitter, the information bits are simultaneously sent over two streams. The first one is assigned to the spatial constellation diagram of size $N_t$, while the second stream is allocated to the signal constellation diagram of size $M$. The bit stream is divided into blocks of $\log_2(N_t M)$ bits. Each block is split into two parallel streams of $\log_2(N_t)$ and $\log_2(M)$ bits, respectively. The former stream activates one of the transmit antennas, and the latter is used to determine a symbol in the signal constellation diagram. More specifically, these $\log_2(N_tM)$ bits are mapped onto a constellation vector $\bf{x}$ of size $N_t$. In SM, only one antenna remains active during transmission, hence only one element in the constellation $\bf{x}$ is non-zero, which is at the position of the active antenna. In particular, given that the antenna index is used as an additional communication resource to transmit information, the active antenna and the symbol index constitute a mapping strategy, and the output vector is written as
\begin{align}\label{eq:1}
{\bf{x}} \triangleq & \Bigl[\begin{array}{ccccccc} \!\! 0 \ & \cdots & \ 0 & \smash{\underbrace{s_m}_{n^{\text {th}} \text { position}}} & 0 \ & \cdots & \ 0 \!\! \\ \end{array}\Bigr]^{T},
\end{align}
where $n=1,2,\ldots,N_t$ represents the active antenna, and $s_\mathit{m}$ is the $m$-th information-bearing symbol from the $M$-ary constellation at the $n$-th position. Fig. \ref{fig:1} schematically depicts the considered system model.

In addition, since the receiver performs energy detection, the amplitude levels in the constellation must be asymmetric in order to avoid having two symbols with the same energy. In consequence, we consider a biased PAM, in which we design non-negative constellation points belonging to the set \cite{li2019channel}
\begin{equation}\label{eq:2}
\mathcal{S}=\left\{0, \sqrt{E_{\rm{min}}},2\sqrt{E_{\rm{min}}}, \ldots, (M-1)\sqrt{E_{\rm{min}}}\right\},
\end{equation}
where $E_{\rm{min}}$ represents the minimum energy difference between adjacent constellation points. Hence, the energy of the $m$-th symbol is $E_m=m^2E_{\rm{min}}$, for $m=0,\ldots, M-1$. Assuming all symbols are equally likely to be transmitted, the average symbol energy is given by \cite{secondini2020direct}
\begin{equation}\label{eq:3}
E_{\rm{av}}=\frac{(M-1)(2M-1)}{6}E_{\rm{min}}.
\end{equation}
Furthermore, at the receiver, the transmitted data from both the activated transmit antenna and the modulation must be detected. Throughout this work, we assume that these two detection processes are independent\footnote{The validity of this assumption lies in the fact that the channel paths are considered uncorrelated, otherwise, the detection processes would be dependent \cite{mesleh2008spatial}.}.
\subsection{Channel Model}
With the focus being on a low-complexity receiver structure, we employ a non-coherent energy detector with knowledge of the fading channel magnitudes \cite{li2019channel}. Specifically, the signal is transmitted over an $N_r \times N_t$ wireless channel $\bf{H}$, and experiences an $N_r$-dimensional additive white Gaussian noise (AWGN) vector ${\bf{w}}=[w_1 \ldots w_{N_r}]^T$. We assume that the signals from all the receive antennas are combined, and the communication channel is quasi-static Rayleigh fading. The $N_rN_t$ entries of $\bf{H}$ are thus independent and identically distributed (i.i.d) random variables, each following a complex circular Gaussian random variable with zero mean and variance $\sigma_h^2$, i.e., having a $\mathcal{CN}(0,\sigma_h^2)$ distribution. In addition, the AWGN vector $\bf{w}$ follows a $\mathcal{CN}(\boldsymbol{0}_{N_r},\sigma_{w}^2{\bf{I}}_{N_r})$ distribution, where $\sigma_{w}^2$ is the variance of each element in $\bf{w}$. We assume a complex baseband signal representation and symbol-by-symbol detection, in which the sampled signal at the $i$-th receive antenna when $s_m$ is transmitted from the $n$-th antenna is written as
\begin{equation}\label{eq:4}
r_i=h_{i,n}s_m+w_i,
\end{equation}
where $h_{i,n}$ is the channel coefficient from the $n$-th transmit antenna to the $i$-th receive antenna, and $s_m$ is the $m$-th information-bearing symbol belonging to the constellation set $\mathcal{S}$.

Moreover, the receiver is assumed to have knowledge of the energy of the signal, hence the output of the energy detector at the $i$-th receive antenna is
\begin{equation} \label{eq:5}
y_i=|r_i|^2=|h_{i,n}s_m+w_i|^2.
\end{equation}
By using energy detector-based receivers, the magnitude of the fading gain from the $n$-th transmit antenna to all received antennas, $|h_{i,n}|$, $i = 1,\ldots, N_r$, can be obtained by transmitting non-zero training symbols, known apriori by the receiver, on the $n$-th transmit antenna while the other antennas remain idle. Since the receiver has only knowledge of the magnitudes of the fading channel gains, then from (\ref{eq:5}), the observation $y_i$ represents a non-central chi-squared random variable with two degrees of freedom, non-centrality parameter $|h_{i,n}s_m|^2$ and variance $\sigma_{w}^2$. By defining $\beta_{i,n}\triangleq |h_{i,n}|^2$, the conditional probability distribution function (PDF) of $y_i$ conditioned on $\beta_{i,n}$ and $E_m$, denoted by $f_{y_i \mid \beta_{i,n},E_m}(y_i \mid \beta_{i,n},E_m)$, is \cite[Eq. (2.44)]{simon2002probability}
\begin{align}\label{eq:6}
f_{y_i \mid \beta_{i,n},E_m}(y_i \mid \beta_{i,n},E_m)=&\frac{1}{2\sigma_w^2}\exp\left(-\frac{y_i+\beta_{i,n}E_m}{2\sigma_w^2}\right) I_0\left(\frac{\sqrt{y_i\beta_{i,n}E_m}}{\sigma_w^2}\right).
\end{align}
We observe from (\ref{eq:6}) that the PDF depends on $\beta_{i,n}$ and $E_m$. Therefore, it is sufficient to consider only the magnitude of the channel coefficients and non-negative, real-valued symbols.

\subsection{ML Decision Rule}
Given that the receiver is equipped with multiple antennas, and due to the independence of the $N_r$ branches, the joint PDF of the received signal is calculated as the product of the marginal PDFs, i.e., 
\begin{equation}\label{eq:7}
f_{{\bf{y}}\mid {\boldsymbol{\beta}}_n,E_m}({\bf{y}} \mid {\boldsymbol{\beta}}_n,E_m)=\prod_{i=1}^{N_r}f_{y_i \mid \beta_{i,n},E_m}(y_i \mid \beta_{i,n},E_m).
\end{equation}
In order to derive the detection metric, the receiver follows the ML decision rule, which jointly searches for ${\boldsymbol{\beta}}_{n}$ and $E_m$ that maximizes the natural logarithm of the likelihood function in (\ref{eq:6}), and is given by \cite[Ch. 7, Sec. 7.3]{simon2005digital}
\begin{align}\label{eq:8}
[\hat{n}, \hat{m}]= \arg \max _{n,m}& \sum_{i=1}^{N_r}\left[\ln \left\{I_{0}\left(\frac{2\sqrt{y_i\beta_{i,n}E_m}}{\sigma_w^2}\right)\right\} - \frac{E_m}{\sigma_w^2}\beta_{i,n}\right],
\end{align}
where $\hat{n}$ and $\hat{m}$ respectively represent the estimated antenna index and symbol. To apply non-coherent detection to low-complexity receivers, we need to consider a simplification of the optimal decision rule in (\ref{eq:8}). One such simplification is based on the asymptotic expansion of the Bessel function $I_0(\cdot)$ and its logarithm which is valid at high SNRs, by considering that $\ln\{I_0\left(|x|\right)\}\approx |x|$, for large $|x|$ \cite{li2019channel}. Therefore, the sub-optimal ML decision rule\footnote{The receiver is sub-optimal due to the simplification of the non-linear Bessel function in (\ref{eq:8}), which makes it applicable to low-complexity receivers \cite{li2019channel}.} becomes
\begin{equation}\label{eq:9}
[\hat{n}, \hat{m}]=\arg \max_{n, m} \sum_{i=1}^{N_r}\left[\sqrt{E_m}\sqrt{y_i\beta_{i,n}}-\frac{E_m}{2}\beta_{i,n}\right].
\end{equation}
In the next section, we present an analytical framework to derive the error rate expressions for the energy-based non-coherent SM detector, defined in (\ref{eq:9}).
\section{Energy-Based Non-coherent SM Detection}
\subsection{Error Rate Performance}
Owing to the distinct signal structure of SM as described in Section II, the transmit vector $\bf{x}$ can be accurately retrieved only when both the transmit antenna index and the transmitted symbol are accurately estimated. Furthermore, the complexity of the detection problem increases significantly when multiple receive antennas are employed. Hence, for analytical tractability, we split the theoretical analysis into two parts, as we show in the sequel. We begin our analysis by focusing on the symbol error rate (SER) $P_{\rm{s}}$, which denotes the probability that the transmitted symbol $s_m$ is received incorrectly. Under equiprobable signaling, $P_{\rm{s}}$ can be expressed as
\begin{equation}\label{eq:10}
P_{\rm{s}}=\frac{1}{M}\sum_{m=0}^{M-1}P_{{\rm{e}}|s_m},
\end{equation}
where $P_{{\rm{e}}|s_m}$ is the probability of error when the symbol $s_m$ is transmitted. Since the decision variable $\bf{y}$ represents a non-central chi-squared random variable with PDF given by (\ref{eq:6}), and by employing the decision rule in (\ref{eq:9}), the probability of detecting $E_{m+1}$ while $E_{m}$ was transmitted refers to the upper tail of the PDF in (\ref{eq:6}), which can be characterized by a chi-squared cumulative distribution function (CDF). By symmetry, the probability of detecting $E_{m-1}$ while $E_{m}$ was transmitted can also be characterized by a chi-squared CDF. Hence, we obtain the following three cases for $P_{{\rm{e}}|s_m}$ depending on the value of $m$
\begin{equation}\label{eq:11}
P_{{\rm{e}}|s_m}=\left\{\begin{array}{lc}
1-F_{\bf{y}}\left(\rho_{m} \mid \beta, E_{m}\right), & m=1, \\
F_{\bf{y}}\left(\rho_{m-1} \mid \beta ,E_{m}\right)+1-F_{\bf{y}}\left(\rho_{m} \mid \beta, E_{m}\right), & 1<m<M, \\
F_{\bf{y}}\left(\rho_{m-1} \mid \beta, E_{m}\right), & m=M,
\end{array}\right.
\end{equation}
where $\rho_m$, $m=0,1,\ldots,M-2,$ is the $m$-th detection threshold, $\beta\!=\!\sum_{i=1}^{N_r} \beta_i$, and $F_{\bf{y}}\left(\rho_{m} \mid \beta, E_{m}\right)$ is the conditional CDF represented as \cite[Eq. (2.45)]{simon2002probability}
\begin{equation}\label{eq:12}
F_{\bf{y}}\left(\rho_{m} \mid \beta, E_{m}\right)= \int_{0}^{\rho_m} f_{{\bf{y}}\mid {\boldsymbol{\beta}},E_m}({\bf{y}} \mid \beta,E_m) {\rm{d}}{\bf{y}} =1-Q_{N_r}\left(\sqrt{\frac{2E_m\beta}{\sigma_w^2}},\sqrt{\frac{2\rho_m\beta}{\sigma_w^2}}\right).
\end{equation}
We note that the detection thresholds $\rho_m$, $m=0,1,\ldots,M-2,$ can be approximated as\footnote{We assume that for high SNRs, the PDFs of the decision variables start to resemble a symmetrical distribution, and the thresholds can be bounded by the midpoint of adjacent symbols' energy levels \cite{anttonen2009error}.}
\begin{equation}\label{eq:13}
\rho_m\approx\frac{\left(E_m+E_{m+1}\right)}{2}.
\end{equation}
Ultimately, by using \eqref{eq:11}, we can reformulate (\ref{eq:10}) as
\begin{align}\label{eq:14}
P_{\rm{s}}=\frac{1}{M}& \sum_{m=0}^{M-2} \mathbb{E}_{\beta}\left\{Q_{N_r}\left(\sqrt{\frac{2E_m\beta}{\sigma_w^2}}, \sqrt{\frac{2 \rho_m \beta}{\sigma_w^2}}\right)+ 1-Q_{N_r}\left(\sqrt{\frac{2E_{m+1}\beta}{\sigma_w^2}}, \sqrt{\frac{2 \rho_m \beta}{\sigma_w^2}} \right) \right\}.
\end{align}
The following proposition gives the final expression for the average SER of the transmitted symbol.

\textit{Proposition 1:} The average SER for detecting the transmit symbol $s_m$ is given by
\begin{align}\label{eq:15}
P_{\rm{s}}=\frac{1}{M}\sum_{m=0}^{M-2}\biggl[\mathcal{A}(E_m) + \mathcal{B}(E_m) + \mathcal{C}(E_{m+1}) \biggr],
\end{align}
where, and $\mathcal{A}(E_m)$, $\mathcal{B}(E_m)$ and $\mathcal{C}(E_{m+1})$ are given by
\begin{equation}\label{eq:16}
\begin{aligned}
\mathcal{A}(E_m)=& \left(\frac{\sigma_w^2}{\delta}\right)^{N_r} \left[{}_2F_1\left(\frac{N_r}{2},\frac{N_r+1}{2};1;\frac{4E_m\rho_m}{\delta^2}\right) + 2 \sum_{k=1}^{N_r-1} \frac{\left(k+N_r-1\right)!}{k! \ (N_r-1)!} \left( \frac{E_m \sqrt{E_m}}{\delta \sqrt{\rho_m}} \right)^{k} \right. \\[3mm]
&\left. \qquad \qquad \qquad \qquad \qquad \quad \qquad \qquad \times {}_2F_1\left(\frac{k+N_r}{2},\frac{k+N_r+1}{2};k+1;\frac{4E_m\rho_m}{\delta^2} \! \right)\right],
\end{aligned}
\end{equation}
\begin{equation}\label{eq:17}
\begin{aligned}
\mathcal{B}(E_m)=&\sum_{n=0}^{\infty} \frac{\left(\rho_m / \sigma_w^2\right)^n}{\Gamma(N_r)}\left(\left( \frac{\rho_m + \sigma_w^2}{\sigma_w^2} \right)^{-n-N_r}\Gamma(n+N_r)-\left(\frac{E_m}{\sigma_w^2}\right)^{-n-N_r}\frac{\Gamma(2n+2N_r)}{\Gamma(n+N_r+1)} \right. \\[3mm]
& \qquad \qquad \qquad \qquad \ \qquad \qquad \left. \times {}_2F_1\left(n+N_r,2n+2N_r;n+N_r+1;-\frac{\rho_m+\sigma_w^2}{E_m}\right) \right),
\end{aligned}
\end{equation}
and
\begin{equation}\label{eq:18}
\begin{aligned}
\mathcal{C}(E_{m+1})=&\sum_{n=0}^{\infty} \frac{\left(E_{m+1} / \sigma_w^2\right)^n}{\Gamma(N_r)}\left(\left( \frac{E_{m+1} + \sigma_w^2}{\sigma_w^2} \right)^{-n-N_r}\Gamma(n+N_r)-\left(\frac{\rho_m}{\sigma_w^2}\right)^{-n-N_r}\frac{\Gamma(2n+2N_r)}{\Gamma(n+N_r+1)} \right. \\[3mm]
& \qquad \qquad \qquad \qquad \qquad \left. \times {}_2F_1\left(n+N_r,2n+2N_r;n+N_r+1;-\frac{E_{m+1}+\sigma_w^2}{\rho_m}\right) \right),
\end{aligned}
\end{equation}
\begin{proof}
See Appendix A.
\end{proof}

Next, we focus on $P_{\rm{a}}$, which is defined as the probability that the antenna index estimate is incorrect. In general, the operation of finding the active transmit antenna is equivalent to solving an $N_t$-hypothesis testing problem at the receiver which involves the computation of multidimensional integrals. Therefore, it is common in the SM literature to compute the average error rate by exploiting union-bound methods. More specifically, the antenna index error rate can be expressed as \cite{di2010space}
\begin{equation}\label{eq:19}
P_{\rm{a}}\leq\frac{1}{2(N_t\!-\!1)}\sum_{u=1}^{N_t}\sum_{v=u+1}^{N_t} P_{\rm{a}}\left(h_{u},h_{v}\right),
\end{equation}
where $P_{\rm{a}}\left(h_{u},h_{v}\right)$ is defined as the probability of error of detecting the index of the active transmit antenna, between antennas $u$ and $v$. In what follows, we focus on calculating $P_{\rm{a}}\left(h_{u},h_{v}\right)$, which is given by
\begin{equation}\label{eq:20}
P_{\rm{a}}(h_u,h_v)=\frac{1}{2}\biggl[ P_{\rm{a}}(h_u,h_v)|_{\hat{n}=u} + P_{\rm{a}}(h_u,h_v)|_{\hat{n}=v} \biggr].
\end{equation}
Let us start by computing $P_{\rm{a}}(h_u,h_v)|_{\hat{n}=u}$. We denote the transmitted symbol energy by $\mathcal{E}$. Furthermore, the squared channel magnitude is denoted by $\beta_n,$ for $n=u,v$. By referring to the ML detector in (\ref{eq:9}), we can write
\begin{equation}\label{eq:21}
P_{\rm{a}}(h_u,h_v)|_{\hat{n}=u}\!=\!\!\left\{\!\!\!\begin{array}{l}
\mathbb{P}\left\{|\sqrt{\mathcal{E}} r|<\frac{\mathcal{E}}{2} (\beta_u+\beta_v)\right\}, \quad \text { if } \!\beta_{u}\! \geq\! \beta_{v}, \\[3mm]
\mathbb{P}\left\{|\sqrt{\mathcal{E}} r|>\frac{\mathcal{E}}{2} (\beta_u+\beta_v)\right\}, \quad \text { if } \!\beta_{u}\!<\!\beta_{v},
\end{array}\right.
\end{equation}
where the sum received signal $r\!\!=\!\!\sum_{i=1}^{N_r}r_i$. Hence, the received signal amplitude $|r|$ can be represented as a Rice distribution with the following CDF \cite[Eq. (2.22)]{simon2002probability}

\begin{equation}\label{eq:22}
F_{|r|}=1-Q_{N_r}\left( \frac{\sqrt{\mathcal{E}\beta_u}}{\sqrt{\sigma_w^2}},\frac{\sqrt{\mathcal{E}\left(\beta_u+\beta_v\right)}}{\sqrt{2\sigma_w^2}} \right).
\vspace{1mm}
\end{equation}
Therefore, $P_{\rm{a}}(h_u,h_v)|_{\hat{n}=u}$ is equivalently written as

\begin{equation}\label{eq:23}
P_{\rm{a}}(h_u,h_v)|_{\hat{n}=u}=\left\{\begin{array}{l}
1-Q_{N_r}\biggl(\sqrt{\frac{\mathcal{E}\beta_u}{\sigma_w^2}},\sqrt{\frac{\mathcal{E}\left(\beta_u+\beta_v\right)}{2\sigma_w^2}} \biggr),\quad \text {\normalfont{ if }} \ \beta_{u} \geq \beta_{v}, \\[6mm]
Q_{N_r}\biggl(\sqrt{\frac{\mathcal{E}\beta_u}{\sigma_w^2}},\sqrt{\frac{\mathcal{E}\left(\beta_u+\beta_v\right)}{2\sigma_w^2}} \biggr), \! \ \quad \quad \text { \normalfont{ if } } \beta_{u}<\beta_{v}.
\end{array}\right.
\vspace{1mm}
\end{equation}
Following similar steps with $P_{\rm{a}}(h_u,h_v)|_{\hat{n}=u}$, we can express $P_{\rm{a}}(h_u,h_v)|_{\hat{n}=v}$ as

\begin{equation}\label{eq:24}
P_{\rm{a}}(h_u,h_v)|_{\hat{n}=v}=\left\{\begin{array}{l}
1-Q_{N_r}\biggl(\sqrt{\frac{\mathcal{E}\beta_v}{\sigma_w^2}},\sqrt{\frac{\mathcal{E}\left(\beta_u+\beta_v\right)}{2\sigma_w^2}} \biggr),\quad \text {\normalfont{ if }} \ \beta_{v} \geq \beta_{u}, \\[6mm]
Q_{N_r}\biggl(\sqrt{\frac{\mathcal{E}\beta_v}{\sigma_w^2}},\sqrt{\frac{\mathcal{E}\left(\beta_u+\beta_v\right)}{2\sigma_w^2}} \biggr), \! \ \quad \quad \text { \normalfont{ if } } \beta_{v}<\beta_{u}.
\end{array}\right.
\vspace{1mm}
\end{equation}
\begin{figure*}[t]
\begin{equation}\label{eq:25}
\begin{aligned}
P_{\rm{a}}(h_u,h_v) \!=&\frac{1}{2}\mathbb{E}_{\beta_u,\beta_v} \! \left\{1\!-\!Q_{N_r}\!\left(\!\frac{\sqrt{\mathcal{E} \beta_{u}}}{\sqrt{\sigma_w^2}}, \! \frac{\sqrt{\mathcal{E}\left(\beta_{u}\!+\!\beta_{v}\right)}}{\sqrt{2\sigma_w^2}}\right)\!+Q_{N_r}\!\left(\!\frac{\sqrt{\mathcal{E} \beta_{v}}}{\sqrt{\sigma_w^2}},\! \frac{\sqrt{\mathcal{E}\left(\beta_{u}\!+\!\beta_{v}\right)}}{\sqrt{2\sigma_w^2}}\right) \! \;\middle|\; \!\beta_{u}\! \geq\! \beta_{v}\right\} \\[4mm]
\!\!&+ \!\! \frac{1}{2}\mathbb{E}_{\beta_u,\beta_v} \!\! \left\{1\!-\! Q_{N_r}\!\left(\!\!\frac{\sqrt{\mathcal{E} \beta_{v}}}{\sqrt{\sigma_w^2}},\! \frac{\sqrt{\mathcal{E}\!\left(\beta_{u}\!+\!\beta_{v}\right)}}{\sqrt{2\sigma_w^2}}\right)\!+Q_{N_r}\!\!\left(\!\!\frac{\sqrt{\mathcal{E} \beta_{u}}}{\sqrt{\sigma_w^2}},\! \frac{\sqrt{\mathcal{E}\!\left(\beta_{u}\!+\!\beta_{v}\right)}}{\sqrt{2\sigma_w^2}}\right)\;\middle|\; \! \beta_{u}\! < \! \beta_{v}\right\}
\end{aligned}
\end{equation}
\hrulefill
\vspace{-2mm}
\end{figure*}
Hence, by combining equations (\ref{eq:23}) and (\ref{eq:24}), we can write (\ref{eq:20}) as given in (\ref{eq:25}) on top of the page. To obtain the average antenna index error rate, we need to evaluate the terms of \eqref{eq:25} by averaging over the channel magnitudes. Upon careful examination, we observe that the computation of the conditional joint expectation over the channel magnitudes for one case is analogous to that of the other. Thus, to avoid redundancy, our attention is directed towards computing the error rate for the scenario where $\beta_{u} \geq \beta_{v}$. The following proposition gives the final form for the average error rate of the antenna index detection.

\textit{Proposition 2:} The average error rate ${P}_{\rm{a}}(h_u,h_v)$ for detecting the index of the active antenna between antennas $u$ and $v$ is given by
\begin{equation} \label{eq:26}
\begin{aligned}
{P}_{\rm{a}}(h_u,h_v)= & \frac{1}{2} \sum_{n=1}^{2} \biggl[ \mathcal{D}_{n} + \mathcal{F}_{n} \biggr],
\end{aligned}
\end{equation}
where $\mathcal{D}_{n}$ and $\mathcal{F}_{n}$, for $n=1,2$, are the average error rate terms for the cases where $\beta_{u} \geq \beta_{v}$ and $\beta_{u} < \beta_{v}$, respectively. The terms $\mathcal{D}_{1}$, $\mathcal{D}_2^{\prime}$, and $\mathcal{D}_2^{\prime \prime}$ are given in \eqref{eq:27}-\eqref{eq:29} on top of the page, where $\mathcal{D}_2 = \mathcal{D}_2^{\prime} + \mathcal{D}_2^{\prime \prime}$.

\begin{figure*}[t]
\begin{equation}\label{eq:27}
\begin{aligned}
\mathcal{D}_{1}=&\sum_{n=0}^{\infty}\int_0^{\infty}\!\!\int_0^{\beta_u}\exp\left(-\frac{\mathcal{E}\beta_u}{2\sigma_w^2}\right)\left(\frac{\mathcal{E}\beta_u}{2\sigma_w^2}\right)^n \frac{\gamma\left(N_r+n,\frac{\mathcal{E}\left(\beta_u+\beta_v\right)}{4\sigma_w^2}\right)}{\Gamma\left(N_r+n\right)}f_{\beta_u}(\beta_u)f_{\beta_v}(\beta_v){\rm{d}}\beta_v {\rm{d}}\beta_u
\end{aligned}
\end{equation}
\begin{equation}\label{eq:28}
\begin{aligned}
\mathcal{D}_{2}^{\prime}=& \sum_{k=1-N_r}^{N_r-1}\!\int_0^{\infty}\int_0^{\beta_u}\exp\left(-\frac{\mathcal{E}(\beta_u+3\beta_v)}{4\sigma_w^2}\right) \left(\frac{2\beta_v}{\beta_u+\beta_v}\!\right)^{\frac{|k|}{2}} I_k\left(\frac{\mathcal{E}\sqrt{\beta_v(\beta_u+\beta_v)}}{\sqrt{2}\sigma_w^2}\right) \\
& \qquad \qquad \qquad \qquad \qquad \qquad \qquad \qquad \qquad \quad  \ \qquad \qquad \qquad \times f_{\beta_u}(\beta_u)f_{\beta_v}(\beta_v){\rm{d}}\beta_v {\rm{d}}\beta_u
\end{aligned}
\end{equation}
\begin{equation}\label{eq:29}
\begin{aligned}
\mathcal{D}_{2}^{\prime \prime}=&\sum_{n=0}^{\infty}\int_0^{\infty}\int_0^{\beta_u}\exp\left(-\frac{\mathcal{E}\left(\beta_u+\beta_v\right)}{4\sigma_w^2}\right)\left(\frac{\mathcal{E}\left(\beta_u+\beta_v\right)}{4\sigma_w^2}\right)^n \frac{\gamma\left(N_r+n,\frac{\mathcal{E}\beta_v}{2\sigma_w^2}\right)}{\Gamma(N_r+n)} \\
& \qquad \qquad \qquad \qquad \qquad \qquad \qquad \qquad \qquad \quad  \ \qquad \qquad \qquad \times f_{\beta_u}(\beta_u)f_{\beta_v}(\beta_v){\rm{d}}\beta_v {\rm{d}}\beta_u
\end{aligned}
\end{equation}
\hrulefill
\end{figure*}
\begin{proof}
See Appendix B.
\end{proof}

Our analysis leads to the conclusion that, in contrast to the calculation of $P_{\rm{a}}$ using conventional union-bound methods, the determination of $P_{\rm{s}}$ is based on the exact error probability associated with the signal constellation diagram. This implies that no union bound is employed in computing this quantity. The exact computation of $P_{\rm{s}}$ avoids the potential inaccuracies inherent in relying on the union-bound for the performance of conventional modulation schemes. Ultimately, since we split our analysis into two parts, i.e., computed $P_{\rm{s}}$ and $P_{\rm{a}}$ separately, we can express the overall average probability of error, $P_{\rm{e}}$, for the energy-based non-coherent SM detector as
\begin{equation}\label{eq:30}
P_{\rm{e}}\leq P_{\rm{a}}+(1-P_{\rm{a}})P_{\rm{s}}=P_{\rm{a}}+P_{\rm{s}}-P_{\rm{a}}P_{\rm{s}},
\end{equation}
where $P_{\rm{s}}$ is given in (\ref{eq:15}) and $P_{\rm{a}}$ is the average antenna index error rate given in (\ref{eq:19}).

\subsection{Diversity Analysis}
In this section, our primary objective is to give insight into the diversity order of the proposed SM energy detector by simplifying the derived expressions. Specifically, the diversity order of the system corresponds to the slope of the error rate curve. In other words, the error rate expression decays as $\gamma_m^{-d}$ at high SNR, where $\gamma_m$ is the SNR of the $m$-th symbol and $d$ is an arbitrary diversity order \cite{wang2003simple}. In our case, the expression we look to evaluate is given in (\ref{eq:30}). Hence, we decide to split our analysis to first find the diversity order of the SER, followed by the antenna index error rate. To tackle this problem, we need to simplify the expressions derived in the previous section. Starting with SER, by defining $\gamma_{m}=E_m / \sigma_w^2$, and $\bar{\rho}_m=\rho_m / \sigma_w^2$, we can rewrite the SER expressions in \eqref{eq:16}-\eqref{eq:18} as
\vspace{1mm}
\begin{equation}\label{eq:31}
\begin{aligned}
\mathcal{A}(\gamma_m)=& \left(1\!+\!\gamma_m\!+\! \bar{\rho}_m\right)^{-N_r} \left[{}_2F_1\left(\!\frac{N_r}{2},\frac{N_r+1}{2};1;\frac{4\gamma_m\bar{\rho}_m}{(1\!+\!\gamma_m\!+\! \bar{\rho}_m)^2}\!\right) \!+\! 2 \sum_{k=1}^{N_r-1} \frac{\left(k+N_r-1\right)!}{k! \ (N_r-1)!} \! \left(\frac{\gamma_m}{\bar{\rho}_m}\right)^{\!k} \right. \\[4mm]
&\left. \qquad \qquad \times \left(\frac{\gamma_m\bar{\rho}_m}{(1+\gamma_m+\bar{\rho}_m)^2}\right)^{\!\frac{k}{2}} {}_2F_1\left(\frac{k+N_r}{2},\frac{k+N_r+1}{2};k+1;\frac{4\gamma_m\bar{\rho}_m}{(1\!+\!\gamma_m\!+\! \bar{\rho}_m)^2} \! \right)\right],
\end{aligned}
\end{equation}
\vspace{1mm}
\begin{equation}\label{eq:32}
\begin{aligned}
\mathcal{B}(\gamma_m)=&\sum_{n=0}^{\infty} \frac{\left(\bar{\rho}_m \right)^n}{\Gamma(N_r)}\left(\left(\bar{\rho}_m + 1 \right)^{-n-N_r}\Gamma(n+N_r)-\left(\gamma_m\right)^{-n-N_r}\frac{\Gamma(2n+2N_r)}{\Gamma(n+N_r+1)} \right. \\[4mm]
& \qquad \qquad \qquad \qquad \quad \ \qquad \qquad \left. \times {}_2F_1\left(n+N_r,2n+2N_r;n+N_r+1;-\frac{\bar{\rho}_m+1}{\gamma_m}\right) \right),
\end{aligned}
\end{equation}
and
\begin{equation}\label{eq:33}
\begin{aligned}
\mathcal{C}(\gamma_{m+1})=&\sum_{n=0}^{\infty} \frac{\left(\gamma_{m+1} / \sigma_w^2\right)^n}{\Gamma(N_r)}\left(\left( \gamma_{m+1} +1 \right)^{-n-N_r}\Gamma(n+N_r)-\left(\bar{\rho}_m\right)^{-n-N_r}\frac{\Gamma(2n+2N_r)}{\Gamma(n+N_r+1)} \right. \\[4mm]
& \qquad \qquad \qquad \qquad \qquad \left. \times {}_2F_1\left(n+N_r,2n+2N_r;n+N_r+1;-\frac{\gamma_{m+1}+1}{\bar{\rho}_m}\right) \right).
\end{aligned}
\end{equation}
In particular, we notice from \eqref{eq:31} and \eqref{eq:33} that that the exponent terms involving the SNR $\gamma_m$ are dominated by $N_r$. Hence, according to \eqref{eq:15}, the average SER achieves a diversity order equal to the number of receive antennas, i.e., achieves full diversity. The following proposition states the diversity order of the antenna index error rate.

\textit{Proposition 3:} At high SNR, the diversity of the antenna index error rate for the non-coherent SM energy detection scheme is proportional to half the number of receive antennas, i.e., $ \propto N_r/2$.
\begin{proof}
See Appendix C.
\end{proof}

Ultimately, by examining the overall error rate expression in (\ref{eq:30}), as the diversity of ${P}_{\rm{s}}$ and ${P}_{\rm{a}}$ are respectively proportional to $\gamma^{-N_r}$, and $\gamma^{-N_r/2}$, we conclude that the SM energy detector achieves a diversity of $N_r/2$, i.e., half the number of receive antennas.

\section{Experimental Validation}
In this section, we aim to analyze the performance of the proposed energy detector for SM via an experimental setup using SDR tools. We start by describing the hardware system setup, then explain the software configurations related to the implementation.
\subsection{Hardware Setup}
\begin{figure*}[!t]
	\centering
	\includegraphics[width=\textwidth]{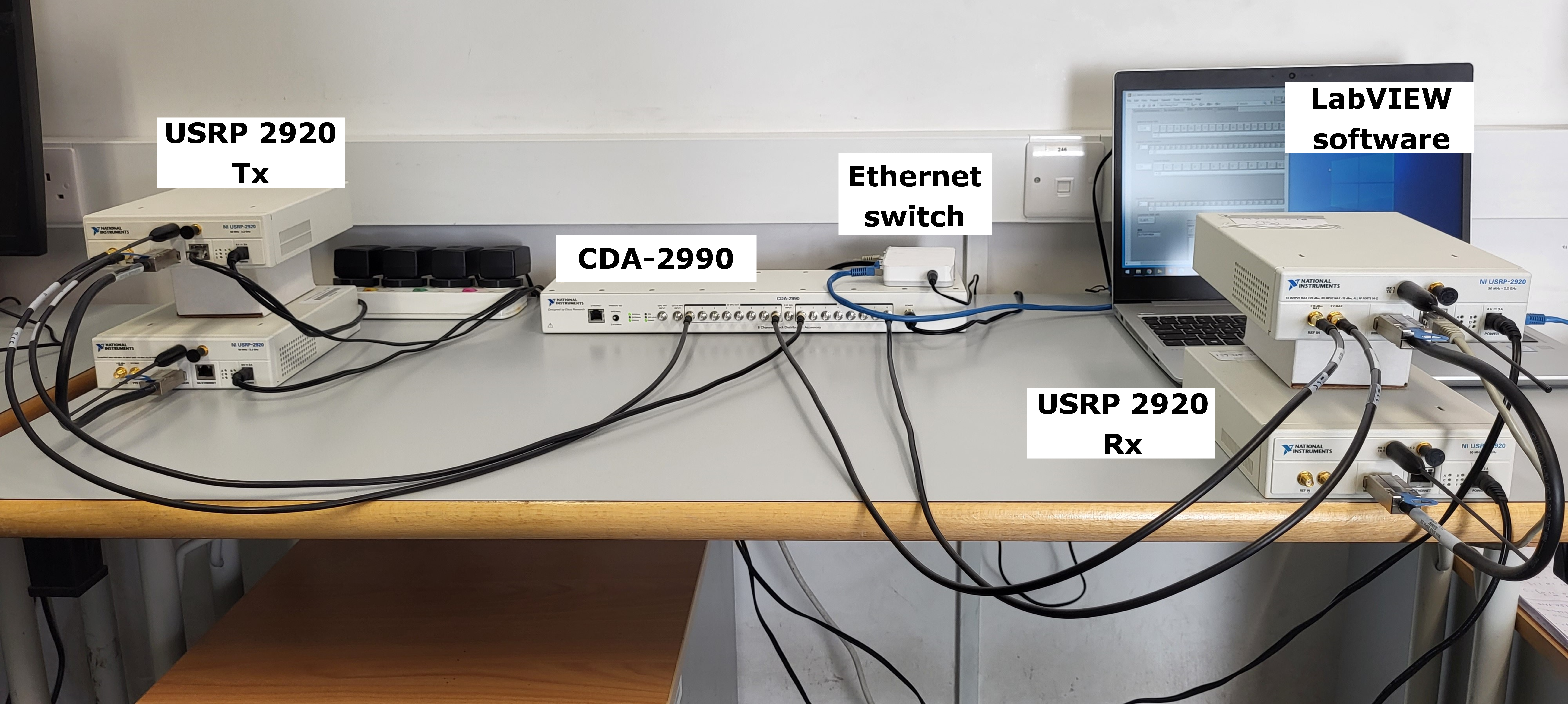}
	\caption{Hardware experimental setup for the realization of the $2\times2$ SM system.}\label{fig:2}
	\vspace{-6mm}
\end{figure*}
We adopt a $2\times2$ SM system configuration, incorporating two transmit and two receive antennas. The rationale behind employing this setup is to introduce and validate the SM energy detection scheme within a simple and tractable framework using the available equipment. More specifically, the hardware consists of four National Instruments Universal Software Radio Peripheral (NI USRP) 2920 devices \cite{ni_USRP}. Each USRP is equipped with a single antenna, as shown in Fig. \ref{fig:2}. The antennas operate at three different frequency bands, i.e., 144 MHz, 400 MHz, and 1.2 GHz; in this experiment, we use the 1.2 GHz band. The transmitter and receiver distance is set to 1 m. Note that the antenna separation at each end of the transmission is equal to $\lambda/2 = 12$ cm, where $\lambda$ is the transmission wavelength, which should result in little to no spatial correlation. Given the physical location and hardware connections, we try to avoid placing the receiver across from the transmitter in an effort to limit the channel to a non-line-of-sight component. The SDR devices are connected to the same network with the use of a Gigabit Ethernet switch. In general, the USRP-2920 devices require an external hardware component to synchronize the transmission. For our specific setup, we employ the CDA-2990 clock distribution module \cite{ni_CDA}, which serves as an external device connecting to both transmitting and receiving ends. Notably, this module is responsible for generating signals at a rate of $10^6$ pulses per second for synchronization. To emulate a MIMO setup, we link a pair of USRP-2920 devices together with the help of a MIMO expansion cable at both sides of the transmission. Moreover, given that every USRP has its own RF chain, we control the activation of the transmit antennas using software which is explained in the next subsection. In this way, we ensure that a single transmit antenna is activated at each transmission time, hence achieving the single-RF chain property of SM. Further details on the hardware equipment are provided in Table \ref{tab:table1}.
\begin{table}[t]
  \begin{center}
    \caption{System setup equipment details.}
    \label{tab:table1}
    \begin{tabular}{|l|l|} 
      \hline
      \textbf{\textbf{Equipment}} & \textbf{\textbf{Description}}\\
      \hline
      Computer & HP Probook 16 GB RAM Core i5 processor\\
      \hline
      SDR & USRP-2920 series, transmit power: 15 to 18 dBm at 1.2 GHz\\
      \hline
      Antennas & VERT400 Tri-band omnidirectional antenna, gain: 3.4 dBi at 1.2 GHz\\
      \hline
      MIMO cable & USRP MIMO data and sync cable\\
      \hline
      Clock distribution & CDA-2990 8-channel clock distribution module\\
      \hline
      Ethernet switch & D-Link Gigabit Ethernet switch\\
      \hline
      Connectors & SMA cables from clock to USRP, Ethernet cables from USRP to switch\\
      \hline
    \end{tabular}
  \end{center}
\end{table}
\subsection{Software Configuration}
\begin{figure*}[!t]
	\centering
	\includegraphics[width=\textwidth]{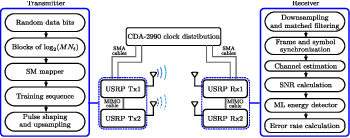}
	\caption{Step-by-step data transmission and reception block diagram with hardware connections.}\label{fig:3}
\end{figure*}

Since our testbed is equipped with NI-USRP 2920 devices, a preferred approach is to conduct signal processing in an offline manner which proves to be more than sufficient to demonstrate our proof-of-concept. Consequently, the software implementation at the transmitter and receiver is realized using the NI LabVIEW programming environment. Specifically, the transmitter generates the information and applies SM to send the data through the wireless channel. Then, the receiver processes the received signal to identify the active transmitting antenna and the transmitted symbol. The block diagram in Fig. \ref{fig:3} depicts the step-by-step signal processing steps at the transmitter and receiver.
\subsubsection{USRP Transmitter}
The transmitter performs the following steps:
\begin{itemize}
\item \textbf{Random bits and framing}: A random bit stream is generated and split into blocks of $\log_2(MN_t)$ bits before passing through the SM mapper.
\item \textbf{SM mapper}: The SM mapper selects the first $\log_2(N_t)$ bits to specify the active transmit antenna, and the remaining $\log_2(M)$ bits are used to choose a symbol from the biased $M$-PAM signal constellation. In this experiment, we have two transmit antennas ($N_t=2$) and choose a biased 4-PAM ($M=4$) modulation.
\item \textbf{Training sequence}: Since one antenna is activated at each transmission time, the training signal is arranged in a similar manner to ensure that the antennas transmit the training symbols sequentially, i.e., one of the two USRP devices transmits the sequence while the other is idle.
\item \textbf{Pulse shaping and upsampling}: A pulse-shaping filter is applied to minimize inter-symbol interference. In this scenario, we apply a root-raised cosine filter with a roll-off factor equal to 0.5. Moreover, each frame is upsampled with an oversampling factor of 12.
\end{itemize}
\subsubsection{USRP Receiver}
The signal received by the USRPs is processed to recover the original data stream. The steps are as follows:
\begin{itemize}
\item \textbf{Downsampling and matched filtering}: The received signal is downsampled by a factor of 12 before applying a matched filter using the same parameters as the transmitter side.
\item \textbf{Synchronization}: This step guarantees that the data gathered by the two receiving antennas are received within the same time frame. More specifically, we implement the Schmidl and Cox algorithm \cite{schmidl1997robust} which searches for the portion of the received signal where two consecutive streams are highly correlated.
\item \textbf{Channel estimation}: The channel estimation operation is achieved by using the least squares algorithm to calculate the channel magnitudes.
\item \textbf{SNR calculation}: Each USRP device has a gain parameter that can be tuned from 0 to 31 dB. Having a higher gain value means having a higher output power. Thus, the process of varying the gains to get an estimate of the SNR is repeated $10^4$ times for every SNR point for the entire transmission frame.
\item \textbf{ML energy detection}: The demodulation process is achieved by applying the ML energy detector in (\ref{eq:9}) to recover the original bit stream.
\item \textbf{Error rate calculation}: The recovered bits from the output of the ML detector together with the SNR measurements are then used to get the average error rate performance for the SM energy detection scheme.
\end{itemize}


\section{Experimental and Simulation Results}
In this section, we present numerical results of the proposed energy detector for non-coherent SM. In the simulations, we consider a flat Rayleigh fading channel with AWGN and channel magnitude knowledge at the receiver. We consider two schemes as benchmarks for performance; C-ML is the SM coherent ML receiver \cite{jeganathan2008spatial}, where $\beta_{i,n}$ is replaced by $h_{i,n}^2$ in (\ref{eq:9}), and MRC is the SM MRC receiver in \cite{mesleh2008spatial}. The proposed receiver is denoted by ED-ML. Ultimately, we evaluate the performance in terms of the average error rate versus the SNR.
\begin{table}
\begin{center}
\caption{Values of the relative error probability versus several summation limits and SNR values. The results are obtained for $M=4$ offset PAM, and $N_t=N_r=2$.}\label{tab:sum_error}
\begin{tabular}{| c | c | c | c |}
 \hline
 $\mathcal{T}$ & \multicolumn{3}{|c|}{$P_e^{\text{rel.}}$}\\
 \hhline{|=|=|=|=|}
 	&	$15$ dB	&	$21$ dB	&	$27$ dB\\
 \hline
 5	&	$0.1282$	&	$0.4113$	&	$0.3$ \\
 \hline
 10	&	$0.0918$	&	$0.2016$	&	$0.1667$ \\
 \hline
 15	&	$0.0459$	&	$0.1129$	&	$0.0667$ \\
 \hline
 20	&	$0.0117$	&	$0.0565$	&	$0.0333$\\
 \hline
\end{tabular}
\end{center}
\end{table}

\begin{figure}
	\centering
	\includegraphics[width=0.67\textwidth]{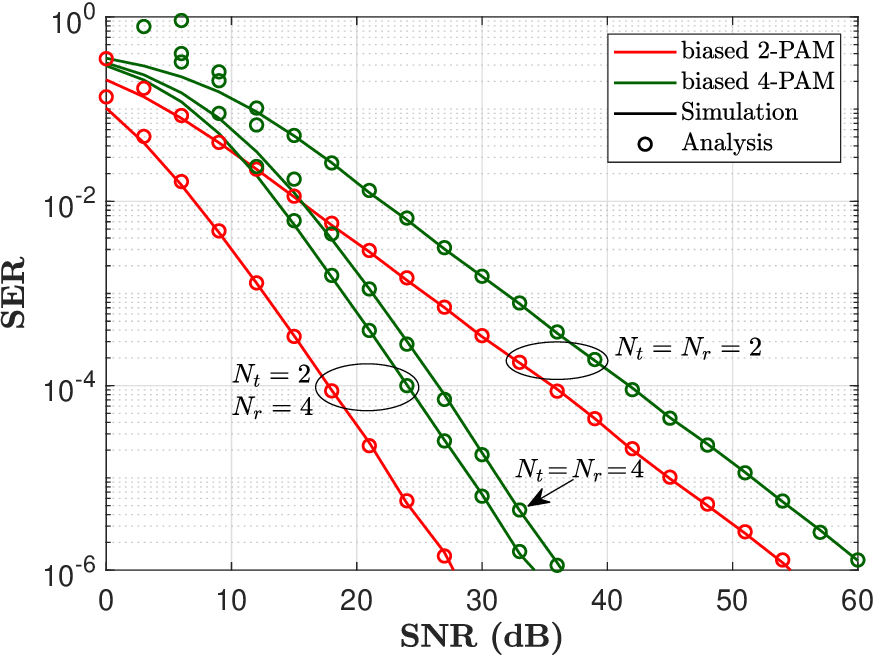}
	\caption{Validation of theoretical analysis for different combinations of system parameters $N_t$, $N_r$, and $M$.}\label{fig:4}
\end{figure}

To quantify the accuracy of the infinite summation, we present in Table \ref{tab:sum_error} values for the relative error between the theoretical and simulated error probabilities, denoted as $P_e^{\text{rel.}}=\left|P_e^{\text{sim.}} - P_e^{\text{ana.}}\right| / P_e^{\text{sim.}}$, for various number of terms $\mathcal{T}$ and different SNRs. The outcomes presented in the table are obtained for $M=4$ and $N_t=N_r=2$. It is evident from the table that the relative error decreases with the increase of the summation terms across all SNR points. Overall, the analytical results are in close agreement with the simulation results, with a relative error $P_e^{\text{rel.}}$ of no more than $0.0565$ for a summation limit of $\mathcal{T}=20$ terms.

We first validate our theoretical error rate analysis in Fig. \ref{fig:4} where we show the performance of the proposed ED-ML receiver for $N_t=2$, $N_r=2,4$, and $M=2,4$. For the analytical results, the summations with infinite limits are truncated to twenty terms. The results demonstrate that there is a perfect match between the numerical and the analytical/theoretical error rate for various cases. We also validate our analysis for the case where $N_t=4$, which shows the tightness of the bound at high SNR for the antenna index detection in (\ref{eq:19}). In addition, we remark that the energy-based SM detector system achieves a diversity order equal to half the number of receive antennas. More specifically, we observe for the case where $N_t=N_r=2$, that the ED-ML receiver achieves a diversity order of $1$. Similarly, for the case where $N_r=4$, the proposed scheme achieves a diversity order of two.

\begin{figure}
\centering
	\includegraphics[width=0.67\textwidth]{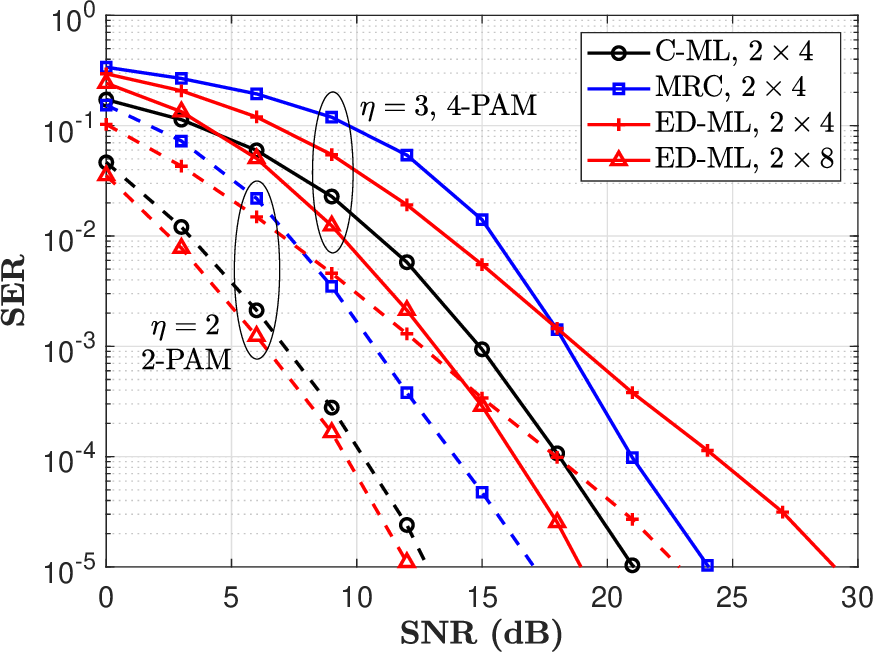}
	\caption{Performance of ED-ML versus C-ML and MRC for $\eta=2, \ 3$ bps/Hz.}\label{fig:5}
\end{figure}

We then compare the performance of ED-ML with that of C-ML and MRC for a spectral efficiency $\eta$ of $2$ and $3$ bps/Hz in Fig. \ref{fig:5}. We first observe that the proposed ED-ML scheme outperforms the MRC in the low SNR regime. The result is based on the fact that MRC requires full channel knowledge to achieve optimal performance, and such a condition is commonly known in the SM literature as constrained channels \cite{jeganathan2008spatial},\cite{maleki2016mrc}. Moreover, we perceive that with the increase in modulation order, non-coherent energy detection outperforms its coherent counterpart. For instance, at an error rate of $10^{-5}$, the gap between biased $2$-PAM (dashed lines) and biased $4$-PAM (solid lines) is around $9$ dB for C-ML, $7.5$ dB for MRC, and $6$ dB for ED-ML. This result highlights the importance of employing a non-negative constellation design which is only optimal for the case of energy-based non-coherent reception. In addition, we plot the performance of the ED-ML scheme for the case where the receiver is equipped with $N_r=8$ antennas (triangle markers). We observe that in order to outperform the C-ML scheme for the same modulation order, the receiver must be equipped with double the number of antennas, hence achieving a diversity order of $N_r / 2$ as was shown in Section III-B. This loss in diversity is due to the absence of full channel knowledge and the fact that the receiver relies on energy detection.

\begin{figure}
\centering
	\includegraphics[width=0.67\textwidth]{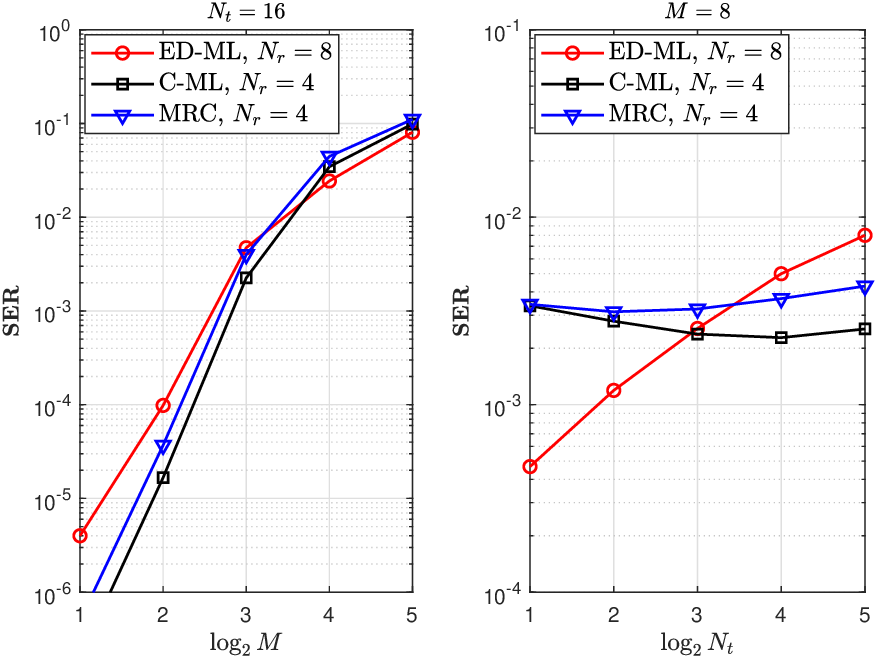}
	\caption{Performance of ED-ML versus C-ML and MRC for different system parameters (SNR $=$ 30 dB).}\label{fig:6}
\vspace{-6mm}
\end{figure}

Furthermore, to quantify the performance of the proposed ED-ML scheme for high spectral efficiencies, we compare its performance with the C-ML and MRC detectors in Fig. \ref{fig:6}. More specifically, we consider two scenarios; the first one where we fix the number of transmit antennas and vary the modulation order, and the second one where we fix the modulation order and vary the number of transmit antennas. We should note that the ED-ML receiver is equipped with twice the number of receive antennas to achieve the same diversity order as the C-ML and MRC receivers. For the first scenario, we observe that the gap between ED-ML and both C-ML and MRC decreases with increasing $M$ until we reach a crossover point after $M=8$. After that, we see that the ED-ML outperforms both detectors. This result highlights again the importance of employing a biased PAM for energy-based non-coherent detection receivers. As for the second scenario, we notice that for a low number of transmit antennas ($N_t \leq 8$), the ED-ML outperforms both detectors. Conversely, for scenarios involving a higher number of transmit antennas, the C-ML detector demonstrates superior performance when compared to the MRC and ED-ML detectors. This result is expected since the receiver jointly detects the transmitted symbol and the index of the active antenna. Thereafter, by employing a biased modulation, we enhance the detection performance, especially when the receiver is restricted to knowledge of the channel gains. However, when the number of transmit antennas increases, the performance is largely affected by the antenna index detection. In summary, we conclude that there is a tradeoff between the number of transmit antennas and modulation order, as we saw that it is preferable to have a higher modulation order and a lower number of transmit antennas for a certain spectral efficiency.


\begin{figure}[t]
\centering
	\includegraphics[width=0.67\textwidth]{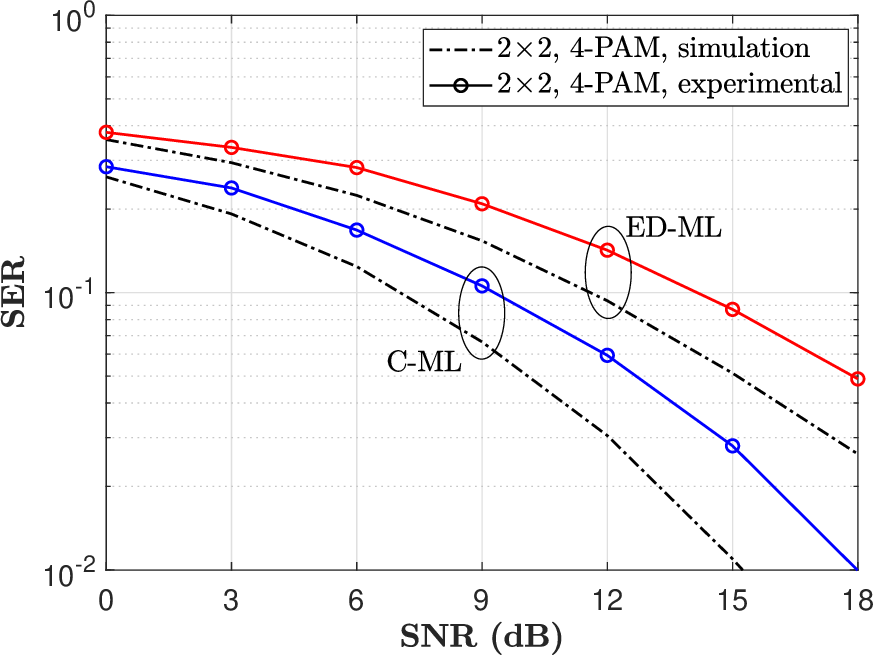}
	\caption{Performance comparison between simulation and experimental results for ED-ML and C-ML, at $\eta=3$ bps/Hz.}\label{fig:7}
	\vspace{-6mm}
\end{figure}

In Fig. \ref{fig:7}, we present the experimental results of our setup of the $2\times 2$ SM system. For the proposed ED-ML scheme, we plot the simulation results, and we show as well experimental results for the case of coherent detection, i.e., the C-ML detector. The experimental results are obtained from the LabVIEW environment by generating a stream of 4000 bits per transmission, and repeating the experiment $10^4$ times for each SNR point in order to obtain good accuracy. Further details on the parameter specifications for the software configuration are in Table \ref{tab:table2}. We notice from Fig. \ref{fig:7} that the maximum measured SNR is 18 dB. One reason for this limitation is that the USRP local oscillator may cause non-constant noise density \cite{mohammadian2021rf}, which affects the SNR measurement. Another reason is having to simultaneously tune the gain parameter of multiple USRP devices to vary the signal power, which changes among devices. Note that other reasons that might contribute to performance degradation exist, however, it is difficult to identify all the imperfections from a hardware perspective.
\begin{table}
  \begin{center}
    \caption{Software parameters settings.}
    \label{tab:table2}
    \begin{tabular}{|l|c|}
      \hline
      \textbf{\textbf{Parameter}} & \textbf{\textbf{Specification}}\\
      \hline
      Operating frequency & 1.2 GHz\\
      \hline
      USRP gain range & 0 to 31 dB\\
      \hline
      Sampling rate & 12 MS/s\\
      \hline
      Modulation & Biased 4-PAM\\
      \hline
      Bits per packet & 4000\\
      \hline
      Training sequence length & 40\\
      \hline
      Oversample factor & 12\\
      \hline
    \end{tabular}
  \end{center}
  \vspace{-8mm}
\end{table}
One solution to this issue is to add a power amplifier to boost the output power, thereby achieving a higher SNR. Nevertheless, we observe from the results that for both schemes, the experimental results follow the same trend as the simulation given the physical environment and hardware constraints, thereby validating the energy detection scheme for SM. We should emphasize that the aim of this experiment is to introduce the use of energy detection in SM systems using the available SDR equipment.
\section{Conclusion}
In this paper, we presented and analyzed an energy-based non-coherent SM detection scheme based on the ML criterion, with channel magnitude knowledge at the receiver. By considering a biased (non-negative) PAM, we studied the performance of the SM energy detector by deriving an accurate analytical framework of the error rate at high SNRs. Moreover, we proved that the diversity order of the proposed SM scheme is equal to half the number of receive antennas. Our results showed that the performance of energy detector-based non-coherent SM is superior to that of its coherent counterpart when employing non-negative constellations. Moreover, we demonstrated in a simple experimental setup that the error rate performance of the proposed energy detection scheme for SM follows the same trend as the simulation results. Finally, due to its low complexity and partial CSI knowledge, an important application for energy-based non-coherent SM is for low-cost low-powered IoT devices.
\appendices
\section{Proof of Proposition 1}
It is important to note that due to the complexity and intractability of the Marcum $Q$-function, it is not straightforward to average the SER over the channel magnitude $\!\beta$. Therefore, we first propose to rewrite the Marcum function in a suitable form to solve the problem at hand, and then obtain the average SER performance at high SNRs. More specifically, we use the following series representation of the generalized Marcum $Q$-function\cite{temme1993asymptotic}

\begin{equation}\label{eq:34}
Q_{N_r}(a,b)=1-\exp\left(\frac{-a^2}{2}\right)\sum_{n=0}^{\infty}\left(\frac{a^2}{2}\right)^n\frac{\gamma\left(N_r+n,\frac{b^2}{2}\right)}{\Gamma(N_r+n)}.
\end{equation}
By applying the following symmetric relation between $Q_{N_r}(a,b)$ and $Q_{N_r}(b,a)$

\begin{equation}\label{eq:35}
Q_{N_r}(a,b)+Q_{N_r}(b,a)=1+\exp\left(-\frac{(a^2+b^2)}{2}\right) \!\! \sum_{k=1-N_r}^{N_r-1} \!\!\left(\frac{a}{b}\right)^k I_k(ab),
\end{equation}
and after a few algebraic manipulations, the first term on the right hand side of (\ref{eq:14}) is given by
\begin{equation}\label{eq:36}
\begin{aligned}
\mathbb{E}_{\beta}\left\{\!Q_{N_r}\!\left(\sqrt{\frac{2E_m\beta}{\sigma_w^2}}, \sqrt{\frac{2 \rho_m \beta}{\sigma_w^2}}\right)\right\}=&\underbrace{\mathbb{E}_{\beta}\left\{\sum_{k=1-N_r}^{N_r-1}\!\!\!\exp\left(\!\!-\frac{(E_m\!+\!\rho_m)\beta}{\sigma_w^2}\right)\!\left(\frac{E_m}{\rho_m}\right)^{\!k} \!I_k\!\left(\frac{2\beta\sqrt{E_m \rho_m}}{\sigma_w^2}\right)\!\right\}}_{\mathcal{A}(E_m)} \\
& \ \ + \underbrace{\mathbb{E}_{\beta}\left\{\sum_{n=0}^{\infty} \exp\left(-\frac{\rho_m\beta}{\sigma_w^2}\right)\left(\frac{\rho_m\beta}{\sigma_w^2}\right)^n\frac{\gamma\left(N_r+n,\frac{E_m\beta}{\sigma_w^2}\right)}{\Gamma(N_r+n)}\right\}}_{\mathcal{B}(E_m)}.
\end{aligned}
\vspace{-2mm}
\end{equation}
Similarly, the second term on the right hand side of (\ref{eq:14}) is
\begin{equation}\label{eq:37}
\begin{aligned}
\mathcal{C}(E_{m+1})\triangleq \mathbb{E}_{\beta}\left\{1-Q_{N_r}\left(\sqrt{\frac{2E_{m+1}\beta}{\sigma_w^2}}, \sqrt{\frac{2 \rho_m \beta}{\sigma_w^2}} \right)\right\} =& \mathbb{E}_{\beta}\left\{ \sum_{n=0}^{\infty} \exp\left(-\frac{E_{m+1}\beta}{\sigma_w^2}\right) \left(\frac{E_{m+1}\beta}{\sigma_w^2}\right)^n \right. \\
& \left. \qquad \qquad \qquad \times \frac{\gamma\left(N_r+n,\frac{\rho_m\beta}{\sigma_w^2}\right)}{\Gamma(N_r+n)} \right\}.
\end{aligned}
\end{equation}
For a conditional error function that is non-zero around the origin and that approaches zero with increasing SNR, the SER is dominated by the near-origin behavior of the PDF of the channel gain and characterizes a high SNR performance\cite{wang2003simple}. Moreover, the Rayleigh distribution of $|h_{i}|$ leads to an exponential distribution of $\beta_i$, since $\beta=\sum_{i=1}^{N_r}\beta_{i,n}=\sum_{i=1}^{N_r}|h_{i,n}|^2$. Therefore, $\beta$ represents a gamma distribution with PDF
\begin{equation*}
f_{\beta}(\beta)=\beta^{(N_r-1)}\exp\left(-\frac{\beta}{(N_r-1)!}\right).
\end{equation*}
Thus, $\mathcal{A}(E_{m})$ in \eqref{eq:36} can be calculated as

\begin{equation} \label{eq:38}
\begin{aligned}
\mathcal{A}(E_{m})=&\sum_{k=1-N_r}^{N_r-1}\int_0^{\infty}\exp\left(-\frac{(E_m+\rho_m)\beta}{\sigma_w^2}\right)\left(\frac{E_m}{\rho_m}\right)^{|k|} I_k\left(\frac{2\beta\sqrt{E_m \rho_m}}{\sigma_w^2}\right)f_{\beta}(\beta){\rm{d}}\beta \\[3mm]
=& \left(\frac{\sigma_w^2}{\delta}\right)^{N_r}\sum_{k=1-N_r}^{N_r-1} \frac{\left(|k|+N_r-1\right)!}{|k|! \ (N_r-1)!} \left( \frac{E_m \sqrt{E_m}}{\delta \sqrt{\rho_m}} \right)^{|k|} \\[3mm]
& \qquad \qquad \qquad \qquad \qquad \qquad \qquad \times {}_2F_1\left(\frac{|k|+N_r}{2},\frac{|k|+N_r+1}{2};|k|+1;\frac{4E_m\rho_m}{\delta^2} \! \right),
\end{aligned}
\end{equation}
where the final form is given in \eqref{eq:16}, and $\delta \triangleq E_m + \rho_m + \sigma_w^2$. Correspondingly, we notice that \eqref{eq:37} and $\mathcal{B}(E_m)$ in \eqref{eq:36} and can be evaluated in a similar manner. Therefore, by using the following integral representation for the lower incomplete Gamma function

\begin{equation*}
\gamma(a,z) = \int_0^z t^{a-1} \exp(-t) {\rm{d}}t,
\end{equation*}
we obtain the final forms of $\mathcal{B}(E_m)$ and $\mathcal{C}(E_{m+1})$ given in \eqref{eq:17} and \eqref{eq:18}, respectively.


\section{Proof of Proposition 2}
As we can see from (\ref{eq:25}), due to the intractability of the Marcum $Q$-function, we follow the same methodology for the computation of the average SER by applying the series representation of the generalized Marcum $Q$-function in (\ref{eq:34}). We can also notice from (\ref{eq:25}) that each error rate expression depends on both channel magnitudes $\beta_u$ and $\beta_v$. Since we consider the case of i.i.d random variables, the fading channel magnitudes are therefore uncorrelated. In consequence, the joint PDF $f_{\beta_u,\beta_v}\left(\beta_u ,\beta_v\right)$ can be written as the product of the marginal PDFs, i.e., $f_{\beta_u,\beta_v}\left(\beta_u ,\beta_v\right)=f_{\beta_u}(\beta_u)f_{\beta_v}(\beta_v)$. Thus, $\mathbb{E}_{\beta_u,\beta_v}\left\{1-Q_{N_r}\left(\frac{\sqrt{\mathcal{E} \beta_{u}}}{\sqrt{\sigma_w^2}}, \frac{\sqrt{\mathcal{E}\left(\beta_{u}+\beta_{v}\right)}}{\sqrt{2\sigma_w^2}}\right)\;\middle|\; \beta_{u}\geq\beta_{v}\right\}$ in (\ref{eq:25}) is given by
\begin{equation}\label{eq:39}
\begin{aligned}
\mathcal{D}_{1}=&\mathbb{E}_{\beta_u,\beta_v}\left\{ \sum_{n=0}^{\infty} \exp\left(-\frac{\mathcal{E}\beta_u}{2\sigma_w^2}\right)\left(\frac{\mathcal{E}\beta_u}{2\sigma_w^2}\right)^n \frac{\gamma\left(N_r+n,\frac{\mathcal{E}\left(\beta_u+\beta_v\right)}{4\sigma_w^2}\right)}{\Gamma\left(N_r+n\right)} \;\middle|\; \beta_{u}\geq\beta_{v}\right\},
\end{aligned}
\vspace{4mm}
\end{equation}
which is rewritten in \eqref{eq:27}. Using (\ref{eq:34}) and (\ref{eq:35}), $\mathbb{E}_{\beta_u,\beta_v}\left\{ Q_{N_r}\left(\frac{\sqrt{\mathcal{E} \beta_{v}}}{\sqrt{\sigma_w^2}}, \frac{\sqrt{\mathcal{E}\left(\beta_{u}+\beta_{v}\right)}}{\sqrt{2\sigma_w^2}}\right)\;\middle|\; \beta_{u}\geq\beta_{v}\right\}$ can be split into two separate expressions given by

\begin{equation}\label{eq:40}
\begin{aligned}
\mathcal{D}_{2}^{\prime}=& \mathbb{E}_{\beta_u,\beta_v}\left\{ \sum_{k=1-N_r}^{N_r-1} \exp\left(-\frac{\mathcal{E}(\beta_u+3\beta_v)}{4\sigma_w^2}\right) \left(\frac{2\beta_v}{\beta_u+\beta_v}\!\right)^{\frac{|k|}{2}} I_k\left(\frac{\mathcal{E}\sqrt{\beta_v(\beta_u+\beta_v)}}{\sqrt{2}\sigma_w^2}\right) \;\middle|\; \beta_{u}\geq\beta_{v}\right\}
\end{aligned},
\end{equation}
and
\begin{equation}\label{eq:41}
\begin{aligned}
\mathcal{D}_{2}^{\prime \prime}=& \mathbb{E}_{\beta_u,\beta_v}\left\{ \sum_{n=0}^{\infty} \exp\left(-\frac{\mathcal{E}\left(\beta_u+\beta_v\right)}{4\sigma_w^2}\right)\left(\frac{\mathcal{E}\left(\beta_u+\beta_v\right)}{4\sigma_w^2}\right)^n \frac{\gamma\left(N_r+n,\frac{\mathcal{E}\beta_v}{2\sigma_w^2}\right)}{\Gamma(N_r+n)} \;\middle|\; \beta_{u}\geq\beta_{v}\right\}.
\end{aligned}
\end{equation}
Therefore, $\mathcal{D}_{2} \triangleq \mathcal{D}_{2}^{\prime} + \mathcal{D}_{2}^{\prime \prime}$. Due to the symmetry of the problem, the error rate expressions for the case where $\beta_u < \beta_v$ can be computed by following the same reasoning as the case of $\beta_u \geq \beta_v$, with the exception that the order of the integrals is inverted. Finally, the antenna index error rate is obtained as given in \eqref{eq:26}.
\section{Proof of Proposition 3}
To prove that the diversity of the antenna index error rate is proportional to $N_r / 2$, it is sufficient to calculate the diversity order of the first term on the right hand side of \eqref{eq:26}, i.e., $\sum_{n=1}^{2} {\mathcal{D}}_{n}$. By using the fact that $\gamma=\mathcal{E}/\sigma_w^2$, we can express ${\mathcal{D}}_{1}$ as
\begin{equation}\label{eq:42}
\begin{aligned}
{\mathcal{D}}_{1}=&\sum_{n=0}^{\infty}\int_0^{\infty}\!\!\int_0^{\beta_u}\exp\!\left(-\frac{\gamma\beta_u}{2}\right)\left(\frac{\gamma\beta_u}{2}\right)^n \frac{\gamma\left(N_r+n,\frac{\gamma}{4}\left(\beta_u+\beta_v\right)\right)}{\Gamma\left(N_r+n\right)}f_{\beta_u}(\beta_u)f_{\beta_v}(\beta_v){\rm{d}}\beta_v {\rm{d}}\beta_u,
\end{aligned}
\vspace{2mm}
\end{equation}
and split $\mathcal{D}_2$ into
\begin{equation}\label{eq:43}
\begin{aligned}
\mathcal{D}_{2}^{\prime}=& \sum_{k=1-N_r}^{N_r-1}\int_0^{\infty}\int_0^{\beta_u}\exp\left(-\frac{\gamma(\beta_u+3\beta_v)}{4}\right) \left(\frac{2\beta_v}{\beta_u+\beta_v}\right)^{\frac{|k|}{2}} \\[3mm]
& \qquad \qquad \qquad \qquad \qquad \qquad \qquad \qquad \times I_k\left(\frac{\gamma\sqrt{2\beta_v(\beta_u+\beta_v)}}{2}\right) f_{\beta_u}(\beta_u)f_{\beta_v}(\beta_v){\rm{d}}\beta_v {\rm{d}}\beta_u,
\end{aligned}
\end{equation}
and
\begin{equation}\label{eq:44}
\begin{aligned}
\mathcal{D}_{2}^{\prime \prime}=&\sum_{n=0}^{\infty}\int_0^{\infty}\int_0^{\beta_u}\exp\left(-\frac{\gamma\left(\beta_u+\beta_v\right)}{4}\right)\left(\frac{\gamma\left(\beta_u+\beta_v\right)}{4}\right)^n\\[3mm]
& \qquad \qquad \qquad \qquad \qquad \qquad \qquad \qquad \qquad \quad \times \frac{\gamma\left(N_r+n,\frac{\gamma\beta_v}{2}\right)}{\Gamma(N_r+n)}f_{\beta_u}(\beta_u)f_{\beta_v}(\beta_v){\rm{d}}\beta_v {\rm{d}}\beta_u.
\end{aligned}
\vspace{2mm}
\end{equation}
Since we are analyzing the diversity of the SM energy detector, we can afford to approximate the derived error rate expressions in order to reveal the system diversity order. By taking this into account, we assume a high SNR regime and start by evaluating $\mathcal{D}_1$. At high SNRs, the lower incomplete gamma function can be approximated by $\gamma\left(a,b\right) \rightarrow \Gamma\left(a\right)$. Using this simplification, we obtain the following closed-form expression (in terms of an infinite sum) for $\mathcal{D}_1$

\begin{equation}\label{eq:45}
\begin{aligned}
\mathcal{D}_1 =& \sum_{n=0}^{\infty}\int_0^{\infty}\int_0^{\beta_u}\exp\left(-\frac{\gamma\beta_u}{2}\right)\left(\frac{\gamma\beta_u}{2}\right)^n f_{\beta_u}(\beta_u)f_{\beta_v}(\beta_v){\rm{d}}\beta_v {\rm{d}}\beta_u  \\[3mm]
=& \sum_{n=0}^{\infty} \left[\frac{2^{N_r}\left( 2+ \gamma \right)^{-n-N_r} \gamma^n \Gamma\left(n+N_r\right)}{\Gamma(N_r)} - \frac{\left(\frac{\gamma}{2}\right)^n \Gamma\left( n+2N_r \right)}{(n+N_r) \Gamma(N_r)^2} \right.  \\[3mm]
&  \left. \qquad \qquad \qquad \qquad \qquad \qquad \qquad \times \ {}_2F_1 \left( n+N_r,n+2N_r;1+n+N_r;-\left(1 + \frac{\gamma}{2}\right) \right) \right].
\end{aligned}
\end{equation}
This result shows that the diversity order of $\mathcal{D}_1$ is proportional to $N_r$, as the exponent terms involving $\gamma$ are dominated by $N_r$. Next, by inspecting (\ref{eq:43}), we notice that the term inside the modified Bessel function (specifically the term under the square root) prevents us from obtaining a significant solution, due to the coupling of $\beta_u$ and $\beta_v$. Consequently, we provide a lower bound for $I_k(\cdot)$, hence a lower bound for $\mathcal{D}_{2}^{\prime}$. Specifically, we use the following bound \cite{baricz2010bounds}
\vspace{1mm}
\begin{equation}\label{eq:46}
I_k(z) > \frac{(z/2)^k}{\Gamma(k+1)}, \text{ for } z > 0, \text{ and } k >-\frac{1}{2}.
\vspace{2mm}
\end{equation}
Therefore, we can rewrite $I_k\left(\frac{\gamma}{2}\sqrt{2\beta_v(\beta_u\!+\!\beta_v)}\right)$ as
\vspace{1mm}
\begin{equation}\label{eq:47}
\begin{aligned}
I_k\left(\frac{\gamma\sqrt{2}}{2}\left(\beta_v \beta_u+\beta_v^2\right)^{\frac{1}{2}}\right) &> \frac{1}{\Gamma(k+1)}\left(\frac{\gamma\sqrt{2}}{4}\left(\beta_v \beta_u+\beta_v^2\right)^{\frac{1}{2}}\right)^k \\[3mm]
&= \frac{\left(\gamma \sqrt{2} / 4\right)^k}{\Gamma(k+1)} \  \sum_{j=0}^{k/2}\binom{k/2}{j} \beta_u^j \beta_v^{k+j/2},
\end{aligned}
\end{equation}
where we used the binomial theorem expansion for $\left(\beta_v \beta_u\!+\!\beta_v^2\right)^{k/2}$, i.e.,
\vspace{2mm}
\begin{equation*}
\left(\beta_v \beta_u+\beta_v^2\right)^{k/2} = \sum_{j=0}^{k/2}\binom{k/2}{j} \left(\beta_v \beta_u\right)^j \beta_v^{k-j/2}.
\vspace{2mm}
\end{equation*}
Then, by using the fact that $I_k(z) = I_{-k}(z)$, we plug the lower bound for the Bessel function in (\ref{eq:43}) and obtain the following lower bound for the expression of $\mathcal{D}_{2}^{\prime}$

\begin{equation}\label{eq:48}
\begin{aligned}
\mathcal{D}_{2}^{\prime} > \sum_{k=1-N_r}^{N_r-1} \sum_{j=0}^{|k|/2} & \left[ \frac{2^{1-|k|}N_r \left(1+\frac{3\gamma}{4}\right)^{-2j-|k|-N_r}\gamma^{|k|} \left(4+\gamma\right)^{-N_r / 2}\binom{|k|/2}{j}}{\Gamma(1+|k|)\Gamma(1+N_r)^2} \right. \\
&\left. \quad \times \left( (4+3\gamma)^{-N_r/2} \Gamma(1+N_r/2)\Gamma(2j+|k|+N_r)\left( 1-(4+\gamma)^{-N_r/2}\right) \right. \right. \\
&\left. \left. \qquad \qquad \qquad \qquad \quad \qquad \times {}_2F_1 \left(\frac{N_r}{2},2j+|k|+N_r;1+\frac{N_r}{2}; -\frac{4+\gamma}{4+3\gamma}\right) \right) \right],
\end{aligned}
\vspace{4mm}
\end{equation}
where we again used the binomial theorem expansion for $\left(\frac{2\beta_v}{\beta_u+\beta_v}\!\right)^{|k|/2}$, i.e.,
\begin{equation*}
\left(\frac{2\beta_v}{\beta_u\!+\!\beta_v}\!\right)^{|k|/2}=(2\beta_v)^{|k|/2} \frac{1}{\sum_{j=0}^{|k|/2}\binom{|k|/2}{j}\beta_u^j \beta_v^{-j + |k|/2}}.
\vspace{2mm}
\end{equation*}
We deduce from (\ref{eq:48}) that the exponent terms involving $\gamma$ are dominated by $N_r/2$. This results states that $\mathcal{D}_{2}^{\prime}$ decays as $\gamma^{-N_r/2}$, hence achieving a diversity order of $N_r/2$. Finally, $\mathcal{D}_{2}^{\prime \prime}$ can be solved in the same manner as $\mathcal{D}_{1}$ by considering the simplification of the lower incomplete gamma function. Thus, an expression for $\mathcal{D}_{2}^{\prime \prime}$ is given by
\vspace{1mm}
\begin{equation}\label{eq:49}
\begin{aligned}
\mathcal{D}_{2}^{\prime \prime} = & \sum_{n=0}^{\infty}\int_0^{\infty}\int_0^{\beta_u}\exp\left(-\frac{\gamma\left(\beta_u+\beta_v\right)}{4}\right)\left(\frac{\gamma\left(\beta_u+\beta_v\right)}{4}\right)^n f_{\beta_u}(\beta_u)f_{\beta_v}(\beta_v){\rm{d}}\beta_v {\rm{d}}\beta_u \\[3mm]
&= \sum_{n=0}^{\infty} \sum_{j=0}^{n} \left[ \frac{2^{-j+2N_r}\gamma^n (4+\gamma)^{-n-N_r}\Gamma(j+N_r)\binom{n}{j}}{\Gamma(N_r)^2} \right.  \\[3mm]
&\left. \qquad \ \ \qquad \times \left( \!\Gamma(-j+n+N_r) - \frac{\Gamma(n+N_r)}{\Gamma(1\!+\!j\!+\!N_r)} {}_2F_1 \left( N_r+j,n+N_r;1+j+N_r;-1 \right)\right) \right],
\end{aligned}
\end{equation}
which achieves a diversity order proportional to $N_r$. We conclude from the analysis of the error rate terms (\ref{eq:45}), (\ref{eq:48}), and (\ref{eq:49}) that the antenna index error rate decays with $\gamma^{-N_r/2}$ at high SNR, which represents the smallest SNR exponent among $\mathcal{D}_{1}$, $\mathcal{D}_{2}^{\prime}$, and $\mathcal{D}_{2}^{\prime \prime}$.
\bibliography{References}
\end{document}